\documentclass[a4paper,fleqn,usenatbib]{mnras}

\usepackage{newtxtext,newtxmath}

\usepackage[T1]{fontenc}
\usepackage{ae,aecompl}

\usepackage{graphicx}	
\usepackage{amsmath}	
\usepackage{amssymb}	
\usepackage{tabu}
\usepackage{booktabs}

\title[Wasp-80b transmission spectroscopy]{Probing the atmosphere of a sub-Jovian planet orbiting a cool dwarf}

\author[E. Sedaghati et al.]{
Elyar Sedaghati$^{1,2,3}$\thanks{E-mail: elyar.sedaghati@dlr.de},
Henri. M. J. Boffin$^3$\thanks{E-mail: hboffin@eso.org},
Laetitia Delrez$^{4,5}$\thanks{E-mail: lcd44@cam.ac.uk},
Micha\"el Gillon$^5$,
\newauthor
Szilard Csizmadia$^1$,
Alexis M. S. Smith$^1$,
Heike Rauer$^{1,2}$
\\
$^1$Institut f\"ur Planetenforschung, Deutsches Zentrum f\"ur Luft- und Raumfahrt, Rutherfordstr. 2, 12489, Berlin, Germany\\
$^2$Zentrum f\"ur Astronomie und Astrophysik, TU Berlin, Hardenbergstr. 36, 10623 Berlin, Germany\\
$^3$European Southern Observatory, Karl-Schwarzschild-Stra$\upbeta$e 2, 85748 Garching bei M\"unchen, Germany\\
$^4$Battcock Centre for Experimental Astrophysics, Cavendish Laboratory, J J Thomson Avenue,Cambridge CB3 0HE, United Kingdom\\
$^5$Institut d'Astrophysique et de G\'eophysique, Universit\'e de Li\'ege, All\'ee du 6 Ao\^ut 19C, 4000 Li\'ege, Belgium\\
}

\date{Accepted 2017 March 10; Received 2017 March 7; in original form 2017 February 13}

\pubyear{2017}

\begin{document}
\label{firstpage}
\pagerange{\pageref{firstpage}--\pageref{lastpage}}
\maketitle

\begin{abstract}
We derive the 0.01 $\mu$m binned transmission spectrum, between 0.74 and 1.0 $\mu$m, of WASP-80b from low resolution spectra obtained with the FORS2 instrument attached to ESO's Very Large Telescope.  The combination of the fact that WASP-80 is an active star, together with instrumental and telluric factors, introduces correlated noise in the observed transit light curves, which we treat quantitatively using Gaussian Processes.  
Comparison of our results together with those from previous studies, to theoretically calculated models reveals an equilibrium temperature in agreement with the previously measured value of 825K, and a sub-solar metallicity, as well as an atmosphere depleted of molecular species with absorption bands in the IR ($\gg 5\sigma$).  Our transmission spectrum alone shows evidence for additional absorption from the potassium core and wing, whereby its presence is detected from analysis of narrow 0.003 $\mu$m bin light curves ($\gg 5\sigma$).  Further observations with visible and near-UV filters will be required to expand this spectrum and provide more in-depth knowledge of the atmosphere.  These detections are only made possible through an instrument-dependent baseline model and a careful analysis of systematics in the data.  ~~~~~~~~~~~~~~~~~~~~~~~ ~~~~~~~~~~~~~~~~~~~ ~~~~~~~~~~~~~~ ~~~~~~ ~~~~~~~ ~~~~~~~~ ~~~~~~~~ ~~~~~~~ ~~~~~~ ~~~~~ ~~~ ~~~~~~~~ ~~~~~~~~~~ ~~~~~~~~~~~~ ~~~~~~~~~~~
\end{abstract}




\section{Introduction}

The remarkable progress made to date in detecting and characterising extra-solar planets has been made possible through dedicated space- and ground-based facilities.  In a mere couple of decades we have gone from discovering the first of these so-called exoplanets \citep{Campbell1988,Wolszczan1992,Mayor1995} to characterising their physical properties, the most fascinating of which has been the detection of their atmospheres.   This is particularly an intriguing feature of these alien worlds, as it provides the means for understanding the mechanisms involved in formation and evolution of planets \citep{Mordasini2012,Mordasini2012b,Dorn2015}, as well as presenting the opportunity for detection of biomarkers \citep{Kaltenegger2009,Snellen2013,Benneke2012}, pointing to perhaps biological processes on those planets capable of harbouring life-forms producing them.

One method through which the atmospheric envelope around a transiting exoplanet is detected is \textit{transmission spectroscopy}, where minute, wavelength-dependent variations of the transit depth are measured from modelling the spectrophotometric transit light curves \citep{Seager2000}.  This method generally probes the upper layers of the atmosphere, where lower pressure levels together with shorter optical paths lead to transmission of stellar radiation through the exo-atmosphere.  This process leaves distinct spectral imprints on the observed radiation, one consequence of which is wavelength-dependent relative planetary radius (obtained from measuring the transit depth).

Such studies are performed with observations either from space \citep{Ehrenreich2007,Sing2011,Deming2013,Wakeford2013,Knutson2014,Sing2016} or with ground-based facilities \citep{Narita2005,Gibson2012gemini,Stevenson2014,Sedaghati2015,Sedaghati2016,Lendl2016,Mallonn2016}, both of which have their own advantages.  Space-based observations have the advantage of not being affected by atmospheric extinction, contamination and turbulence, and therefore benefit from having the entire electromagnetic range as probing domain.  For instance, NASA's \textit{James Webb Space Telescope} \citep[JWST;][]{Gardner2006} will be able to probe exo-atmospheres in the infrared wavelengths, where transmission spectroscopy from the ground is rather difficult.  On the other hand, ground-based facilities benefit from telescopes with large collecting mirrors, which is essential for performing spectroscopy at high time- and spectral resolution.

ESO's FOcal Reducer and low dispersion Spectrograph \citep[FORS2;][]{Appenzeller1998} mounted at the Cassegrain focus of Unit Telescope 1 (UT1) of the Very Large Telescope (VLT) has been the instrument of choice to perform such observations \citep{Bean2010,Bean2011,Sedaghati2015,Sedaghati2016,Lendl2016,Nikolov2016}.  It offers two multi-object spectroscopic modes: (1) Mask eXchange Unit (MXU, custom designed and laser-cut masks), (2) MOS (Multi-Object Spectroscopy, movable slitlets via 19 pairs of arms).  The option of wide slits, multiple reference star selection for telluric correction in conjunction with the great light collecting power of the 8.2m telescope, make this instrument ideal for performing differential spectrophotometric observations, required here.

Previously, it was determined that the Longitudinal\footnote{Or Linear as both adjectives are used in the definition of the LADC.} Atmospheric Dispersion Corrector \citep[LADC;][]{Avila1997} of the FORS2 instrument, a pair of movable prisms, introduced systematic effects in exoplanet transit light curve observations due to the degraded anti-reflective coating causing differential transmission through the optics \citep{Boffin2015}.  This problem was subsequently ratified by the upgrade of the LADC unit and the improvements were initially highlighted by \cite{Sedaghati2015}.  In this study we present results from observations of WASP-80, a dwarf star with a transiting gas giant, that were taken prior to the aforementioned upgrade.  Therefore, we will consider possible implications of the systematic effects introduced by the degraded LADC unit in the following analysis.

WASP-80b is a gas giant transiting a cool, possibly late K-type 11.88V magnitude dwarf, with a mass of 0.554 $\pm$ 0.035 M$_{\mathrm{jup}}$ and radius of 0.952 $\pm$ 0.026 R$_{\mathrm{jup}}$, orbiting its host star with a 3.068 day period \citep{Triaud2013}.  It is one of only a handful of gas giants orbiting a late-type dwarf host \citep[e.g. WASP-43b, Kepler-45b, HAT-P-54b, HATS-6b;][]{Hellier2011,Johnson2012,Bakos2014,Hartman2015} and has a day-side temperature within the T-dwarf range \citep{Triaud2015}.  These authors also found a transmission spectrum indistinguishable from a flat line and no evidence of active region crossing by the planet \citep{Triaud2013,Mancini2014}, despite the stellar spectral analysis indicating high levels of activity \citep{Mancini2014}.

In this work, we present a red transmission spectrum of this planet from light curves that are somewhat compromised by systematic effects.  We will study the role of telluric and instrumental effects in introducing such discrepancies in the light curves, before consideration of any astrophysical phenomena.

\begin{figure}
\includegraphics[width=\linewidth]{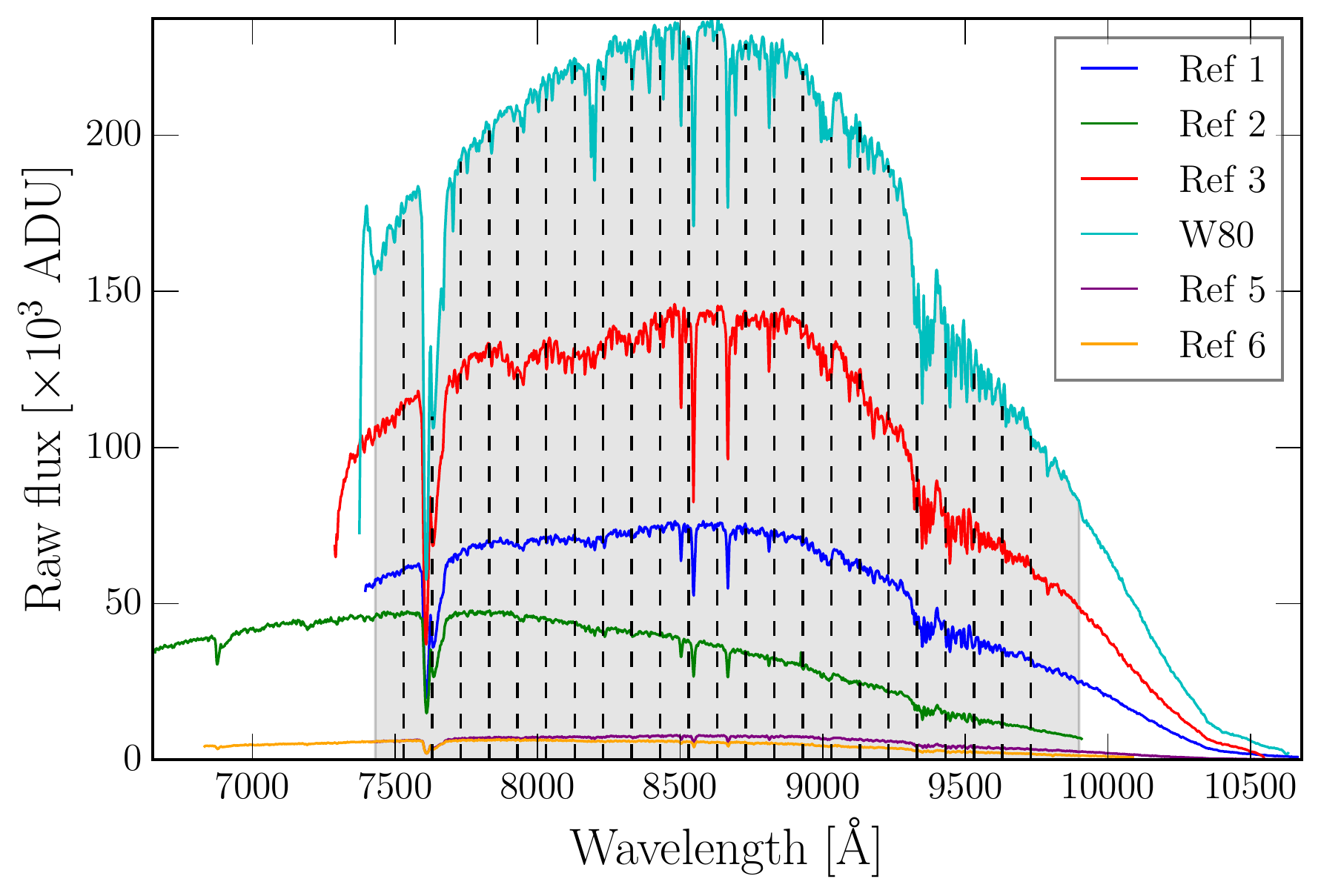}
\caption{Wavelength calibrated and corrected spectra of all the stars observed with the MXU.  The grey area shows the wavelength domain used to obtain all the broadband light curves.  The dashed lines show the boundaries of the spectrophotometric channels, used for integration of WASP-80 and reference stars.}
\label{fig:Raw Spec}
\end{figure}

\section{Data analysis}\label{sec:Data analysis}
\subsection{Observations}

A single transit of WASP-80b was observed using Antu, the 8.2m UT1 of the VLT with the FORS2 instrument on the 16th of June 2013, with the data taken as part of the programme 091.C-0377(C) (PI: Gillon), before the LADC upgrade.  The programme itself comprised four transit observations, two of which were not performed due to bad weather, while the light curves from the fourth transit suffered from severe systematic effects and are subsequently not used for the analysis in this work.  The instrument has a 6.8$\arcmin \times 6.8\arcmin$~field of view and is equipped with two detectors, where the red-optimised MIT set was utilised for the observations.  The instrument was used in the MXU mode, which means that a custom designed mask with 10\arcsec-wide slits positioned on the target and comparison stars, was placed in the light path.   This essentially acts as a blocking mask to free the CCD for recording of the simultaneous spectra.  The 600z grism (with the order sorter filter OG590) was used as the dispersing element, yielding a spectral range of $\sim 0.73-1.05~\mu$m, although the exact wavelength coverage for each target is dependent on the horizontal location of its slit on the CCD.

The entire science observational sequence lasted $\sim$4.87h, with the first frame taken at 05:17UT and the last taken at 10:04UT.   The complete transit lasted $\sim$2.26h, with first contact at 06:17UT and the last one at 8:32UT.  The LADC was left in park position during the observing sequence, with the two prisms fixed at their minimal separation distance of 30mm.  The standard 100kHz readout mode was used, which together with exposure time of 25s (apart from the first seven frames where adjustments were made in order to reach the optimal value), yielded 277 exposures, 130 of which were during the transit.  The conditions were clear throughout the night and the seeing varied between 0.97\arcsec~and 2.35\arcsec.  The field started at airmass of 1.24, rose to 1.08 and the last frame was taken at airmass 1.45.  A copy of the mask was created for the purpose of wavelength calibration, with narrow 1\arcsec~slits centred on the science slits. Bias, flatfield and arc-lamp (for the purpose of wavelength calibration) images were taken, as part of the routine daytime calibration sequences, before and after the observations.

\begin{figure*}
\includegraphics[width=\textwidth]{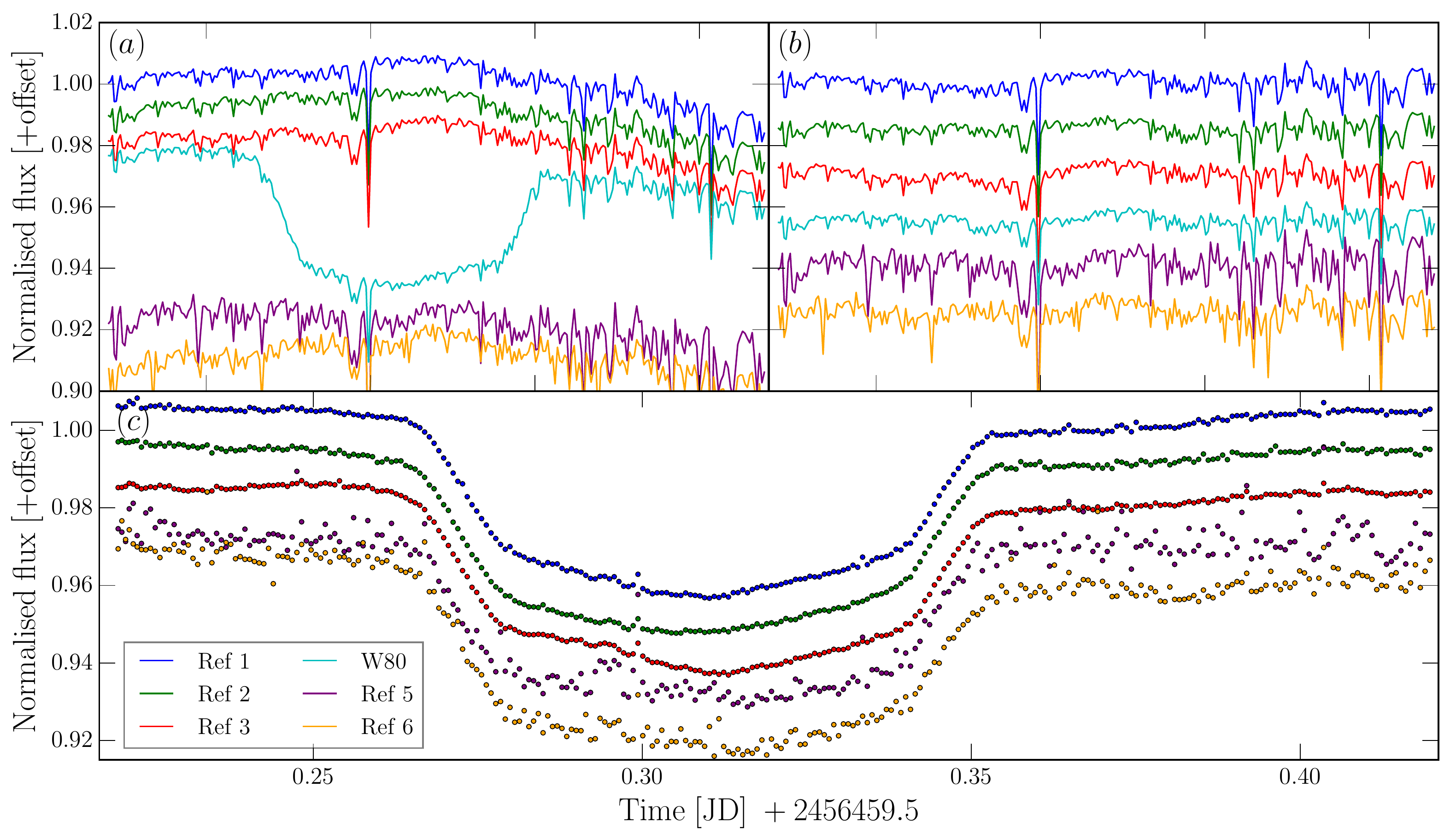}
\caption{(\textit{a}) Raw broadband light curves, normalised to out of transit level and shifted for visual aid.  (\textit{b}) Broadband light curves corrected for airmass.  The transit has been removed from the WASP-80 data.  This data is used to check for correlations with other physical parameters, besides the airmass.  (\textit{c})  Differential transit light curves, again normalised to the out of transit flux level and shifted for clarity.  The colours correspond to the raw light curves.  These data are not corrected for the airmass dependent variations, as this will be part of the analytical transit model.}
\label{fig:Raw LCs}
\end{figure*}

\subsection{Data reduction}

We used our previously written reduction pipeline based on {\tt PyRAF} \citep{Sedaghati2016}, which is optimised to the science goals of transmission spectroscopy.  In addition to the standard steps of overscan and bias shape subtraction, spectral flat-fielding and wavelength calibration, sky background subtraction and cosmic ray contamination removal, the pipeline also includes steps for optimizing the size of the extraction box used to obtain the one-dimensional stellar spectra, performed independently for each star.  We also experiment with the optimal extraction algorithm \citep{Horne1986} and check to see if this in any way improves the quality of the extracted spectra, which was not the case.  Once the wavelength calibrated spectra  are obtained, the pipeline makes corrections to the dispersion solution calculated for each target.  This is first done for all the spectra of each star through optimisation of the cross correlation function, and then the spectra of the stars are further corrected with respect to each other to ensure a consistent solution across all apertures.  This is a necessary and important step due to the low resolution of the spectra.  An example set of these wavelength calibrated and corrected spectra is shown in Figure \ref{fig:Raw Spec}.

\subsection{Broadband light curves \& systematic effects}

To obtain the broadband light curves, we integrate the series of spectra for each star within the largest possible common domain (shown as the grey area in Figure \ref{fig:Raw Spec}), which are shown in Figure \ref{fig:Raw LCs}(a).  At first glance we observe an airmass-dependent trend for all the targets, as well as the clear transit signal of WASP-80b.   Initially we correct all the light curves for both of these factors, the results of which are shown in Figure \ref{fig:Raw LCs}(b).  These values are later on used to search for possible physical variants responsible for some of the systematic trends in the final differential transit light curve.  We also produce differential transit light curves of WASP-80 with respect to all the observed comparison stars, shown in Figure \ref{fig:Raw LCs}(c).  This shows a transit light curve that is very precise and its deviations from a transit model are mostly due to systematic trends.  Most significantly, we observe an almost V-shape transit, which could be interpreted as crossing of a stellar spot by the exoplanet, as was observed for WASP-52b \citep{Kirk2016}.  However, this claim cannot be made until all possible sources of correlated noise in the data have been considered.

\subsubsection{Instrumental effects}

Prior to the upgrade of the FORS2 instrument, where the LADC prisms \citep{Boffin2015} were replaced, inhomogeneities in the degraded anti-reflective coating of the old unit were causing differential transmission through the telescope optics.  This would then manifest itself as systematic flux variations in the differential light curves.  We now take a closer look at this effect.

\begin{figure}
\includegraphics[width=\linewidth]{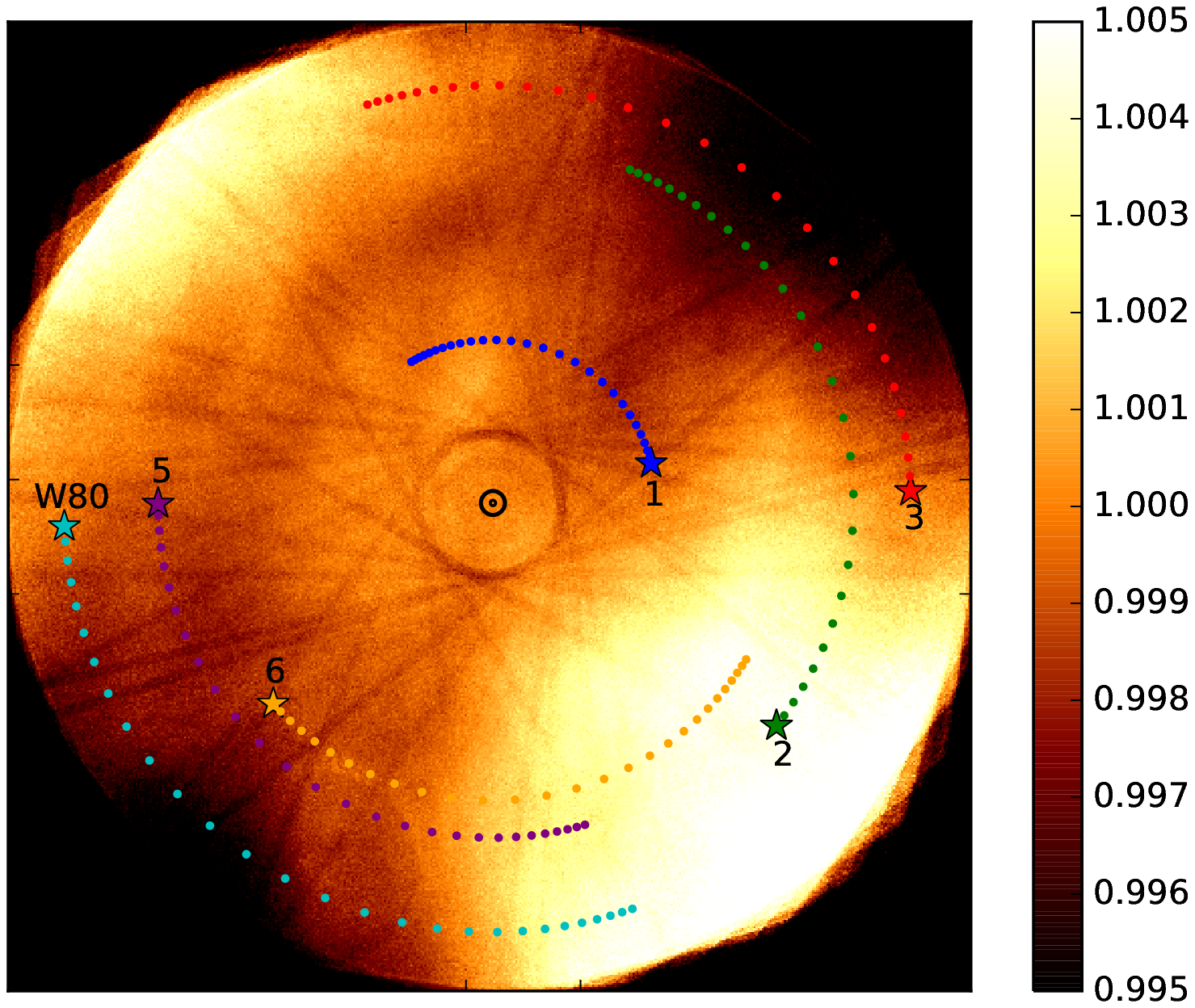}
\caption{Studying the systematic effects due to the old degraded LADC prisms from the stacked flat-field image.  The colour bar on the right shows the scale of the effect (i.e. between $-0.5\%$ and $+0.5\%$). The plotted stars are the 6 observed targets at the starting position that optimises the correlation with the light curve residuals, calculated from Figure \ref{fig:LADC corr}.  The dots represent the position of the star in every 10th frame.  The colours are consistent with all the light curve plots and additionally each star has been labeled corresponding to the aperture number, used consistently throughout the paper.}
\label{fig:LADC}
\end{figure}

\cite{Boffin2016} presented the improvement in the transmission of light through the optical elements of FORS2 by comparing a stack of flatfield images from before and after the LADC exchange.  For our analysis of the systematic trends attributed to the old degraded LADC (used for the observations here), we use their stacked flatfield image taken with the R\_SPECIAL+76 filter, shown in Figure \ref{fig:LADC}.  The individual twilight flatfields used to construct this frame were taken approximately a year after our observations.  To this effect, the exact initial position of the field of view relative to the LADC configuration, inferred through observation of flatfield inhomogeneities, is not precisely known.  Therefore, we can only try to estimate this position from possible trends in the light curves.  In Figure \ref{fig:LADC} we plot the observed stars at this estimated starting position and trace the path of each star through the entire observing sequence, where every 10th exposure is shown with a dot.

\begin{figure}
\includegraphics[width=\linewidth]{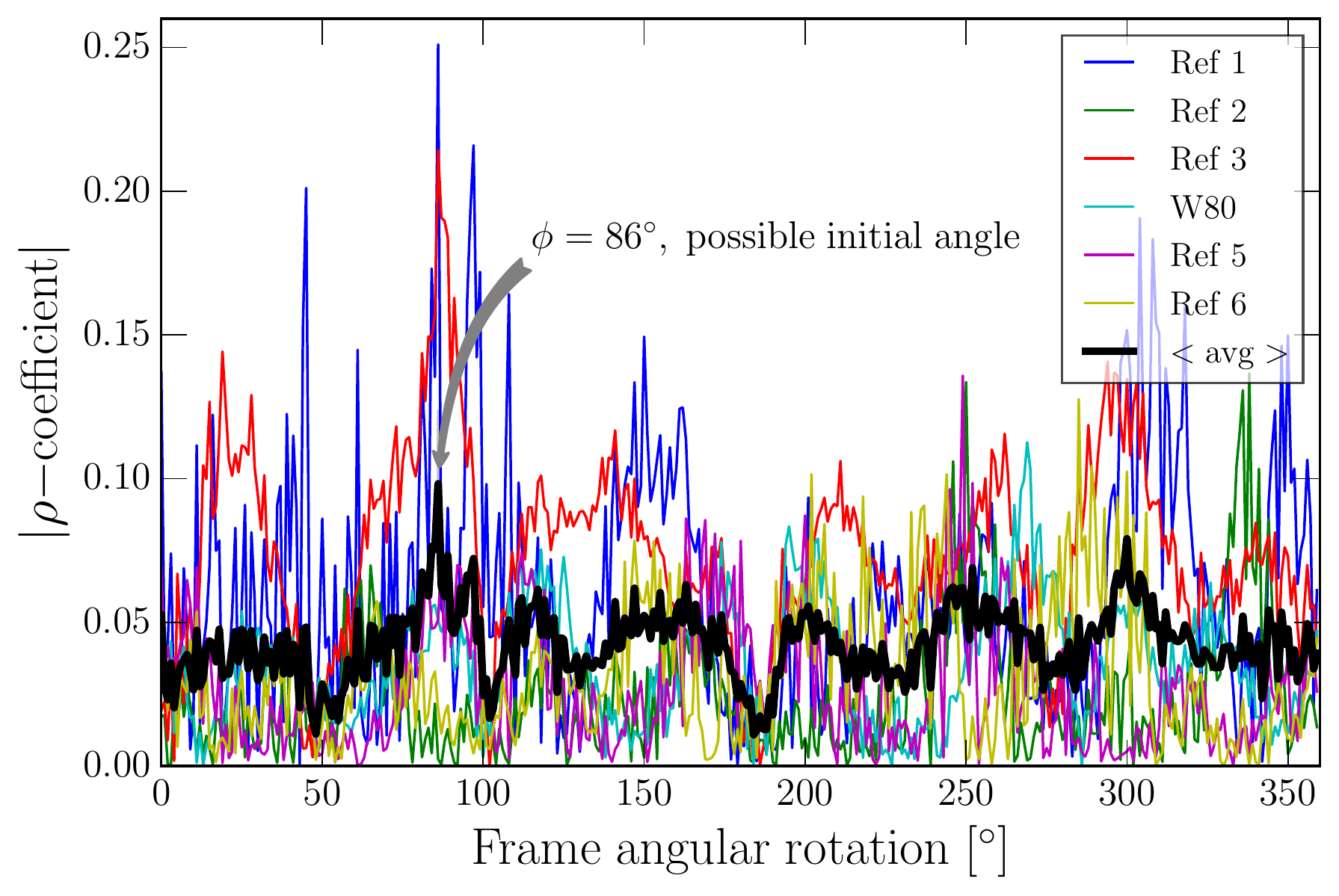}
\caption{Determination of LADC deficiencies and light curve systematic trends.  The initial setup shown in Figure \ref{fig:LADC} is rotated, and the correlation between the flatfield value along each path, with the corresponding light curve is calculated.  The thick black line represents the mean correlation value for all the stars.  A lack of strong correlation indicates that the degraded LADC is not a significant cause of systematic trends in the light curves.}
\label{fig:LADC corr}
\end{figure}

\begin{figure}
\includegraphics[width=\linewidth]{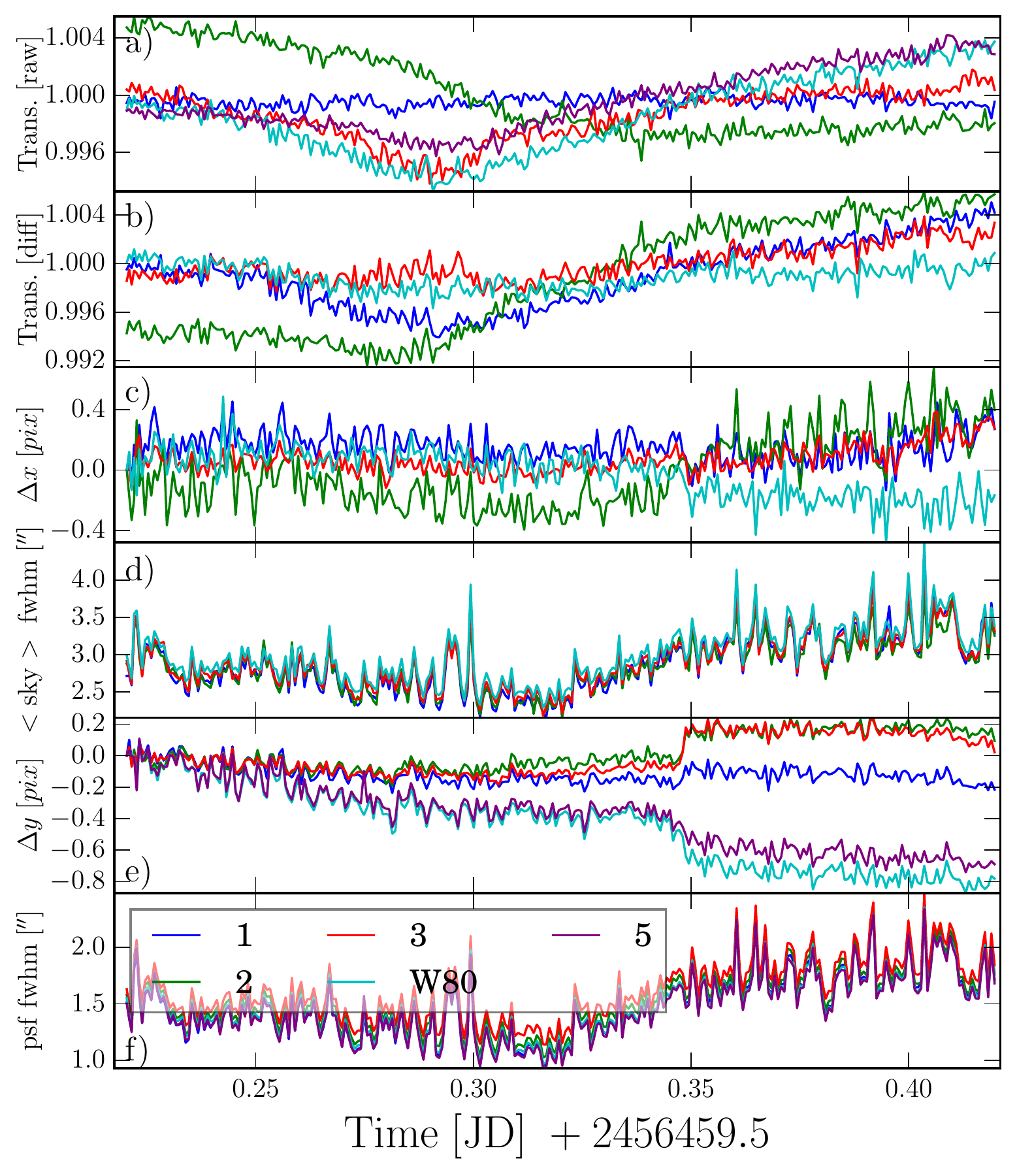}
\caption{Measurement of physical variants that could possibly introduce systematic trends in the light curves. (a) Shows the measurement of normalised flatfield flux along the (most likely) path of the individual stars, and (b) represents the same measurements but relative to the values of WASP-80's path.  (c) Shows the measurement of spectral shift in the dispersion direction relative to the first observation calculated by measuring the centres of multiple telluric absorption features across the entire spectrum, and (d) is the average FWHM values of those fitted gaussian profiles, which is dependent on seeing conditions.  (e) Indicates the drift of spectra along the physical axis relative to the first frame, calculated by again fitting multiple gaussian functions along the physical axis of the two dimensional spectra, and (f) shows the variations of FWHM values of those fitted profiles.  These variations are identical in shape to what was measured along the dispersion axis as expected, and are used as a direct measure of seeing condition variations.  These parameters have not been calculated for reference star 6 due to its very low spectral SN.}
\label{fig:optical state}
\end{figure}

To obtain the possible starting point of the field of view relative to the LADC setup, we rotate the configuration shown in Figure \ref{fig:LADC} for 360$^\circ$ at 1$^\circ$ steps to cover all possible initial positions.  If the LADC deficiencies are the cause of systematic trends in the light curve, then at the true path of the field there will be a spike in the calculated correlation between the light curve of each star (the ones that are corrected for airmass and transit, ref. Fig \ref{fig:Raw LCs}(b)) and the flatfield value along the path taken by the star, shown in \ref{fig:LADC} and read in \ref{fig:optical state}(a,b).  This calculated correlation as a function of field rotation angle is shown in Figure \ref{fig:LADC corr}, where the indicated peak at 86$^\circ$ starting angle could point to a possible relative orientation of the LADC at the beginning of the observing sequence\footnote{It must be noted that this value is 3$^{\circ}$ offset from what is calculated based on the orientation of the stacked flatfield image.}.  As a measure of correlation, we simply choose the \textit{Pearson's correlation coefficient}, $\rho$, defined as the ratio of the covariance of the two variables ($\mathrm{cov}(x,y)$) to the product of their respective standard deviations ($\sigma_x,\sigma_y$).

In order to study the impact of the LADC inhomogeneities on the measured light curves, we read the normalised flatfield value along the path of each star, for this given starting position, which is shown in Figure \ref{fig:optical state}(a).  We use the normalised flatfield, as its variations are a direct indication of the problems caused by the degraded LADC coating.  Additionally, since we are also interested in the correction of the differential transit light curves, as a second order correction, we also calculate the ratio of these readings to the values read along the path of WASP-80.  This is shown in Figure \ref{fig:optical state}(b).  The calculated correlation between the light curve residuals and the flatfield variations due to the degraded LADC is rather weak, which means that the LADC is not a major contributor to the systematic trends.  However, it must be noted that the flatfield images used to construct the composite image in Figure \ref{fig:LADC} were taken more than a year after the epoch of the observations analysed in this work.  Therefore, our conclusion about the impact of the LADC deformities is rather tentative. 

A further possible cause of correlated noise is the spatial stability of the instrument.  We characterise the positioning of stellar spectra on the detectors along two axes, spatial and spectral ($\Delta y$ \& $\Delta x$).  The former variation is measured by taking the mean of the mid-points of several fitted Gaussian profiles to the two-dimensional images across the entire frame in the spatial direction, which are plotted relative to the initial frame in Figure \ref{fig:optical state}(e).  We measure the latter in a similar way, but instead fit several telluric absorption features in the one-dimensional spectra, shown in Figure \ref{fig:optical state}(c).  Studying the relations between both of these variables indicated no significant correlation with the flux variations, which is perhaps due to the stability of the instrument and the data reduction procedure accounting for any remaining residual effects.

\begin{figure}
\includegraphics[width=\linewidth]{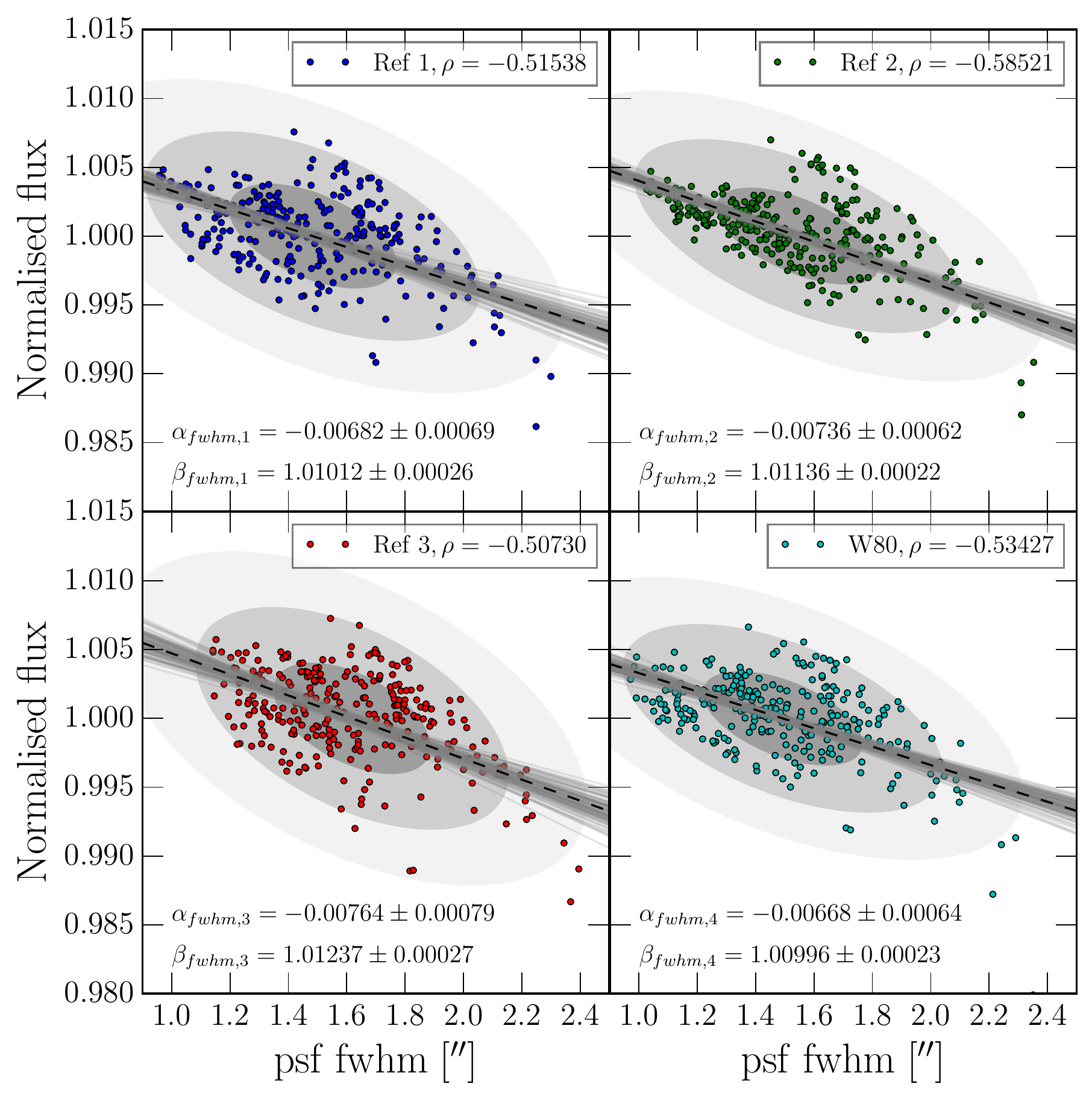}
\caption{Correlation of seeing variations and the raw stellar light curves, for the four brightest targets observed.  The Pearson's correlation coefficient, $\rho$, which shows a consistent and moderate correlation, has been given for each star.  This correlation relation is further demonstrated through the 1-, 2- and 3-$\sigma$ ellipsoids drawn with different shades of grey. In order to associate the seeing values with flux variations, we fit the mean-centred data using an orthogonal linear regression model \citep{Isobe1990} shown as black dashed line, whose parameters are given for each target.}
\label{fig:seeing correlation}
\end{figure}

\subsubsection{Telluric effects}

In addition to the instrumental effects mentioned previously, telluric conditions also play a significant role in introducing time-correlated noise into the measured time-series data.  Namely, variations in the measure of atmospheric seeing, due to turbulence, cause changes in the stellar point spread function (PSF) along both the dispersion and the physical axes.  We measure the former as the FWHM of Gaussian profile fits to several telluric absorption lines in the one dimensional spectra, variations of which are shown in Figure \ref{fig:optical state}(d) and denoted as ``<sky> fwhm''.  This is essentially the convolution of the instrumental profile and the true stellar profile, and therefore a measure of variations in the stellar psf given the instrument response is constant.  The FWHM of the PSF along the physical axis is also measured through recording the characteristics of several Gaussian function fits to various columns (orthogonal to the dispersion axis) of the two dimensional spectra.  These are shown in Figure \ref{fig:optical state}(f), as ``psf fwhm'' in units of arcseconds, which is another approach to measuring the seeing variations.

We compare the variations of seeing measured separately for each star (from the psf FWHM) to the normalised flux after the correction for the airmass trend.  For WASP-80, additionally, the transit feature is also removed for these calculations, in order to maintain a constant degree of freedom in all correlation plots.  Such relations are plotted in Figure \ref{fig:seeing correlation}, where the comparison stars 5 and 6 have not been included due to their significantly low SN.  This will be the case for the remainder of our analysis.

For the measure of correlation, we again use Pearson's parameter described briefly earlier, the calculated values for which are given in each panel.  Additionally, we calculate the 1-, 2- and 3-$\sigma$ confidence intervals, shown as different shades of grey for visualization purposes.  The levels of correlation between individual flux variations and atmospheric seeing are consistent across all the stars and point to a \textit{moderate} association between the two parameters.  The relation between each pair of parameters is calculated using an \textit{orthogonal least squares} approach \citep{Isobe1990}, using the {\tt odr} package from {\tt python}'s {\tt scipy} \citep{scipy}.  This is preferred to the standard linear regression approach due to the present uncertainties in determining ``psf FWHM''.  We use the gradients of these relationships, $\alpha_{\mathrm{fwhm}}$ given in Figure \ref{fig:seeing correlation}, to transform the measured seeing values to flux variations for each star, which in turn will be used as an input of our systematic model, to be described shortly.  We choose this approach instead of constructing a correction model to avoid the introduction of biases in our parameter determination and error estimation.

\subsection{Light curve models}

We model the transit light curves with the implementation of the analytic solution of \cite{Mandel2002} as the mean function.  Additionally, we estimate the correlated noise as a Gaussian Process (GP), which provides a model-independent stochastic method for the inclusion of the systematic component into our model \citep{Rasmussen2006}.  The application of GPs to modelling exoplanet transit light curves and transmission spectroscopy was introduced by \cite{Gibson2012GPs}.  The strength of this approach in accounting for systematic trends was highlighted in the detection of hazes in the atmosphere of the exoplanet HD189733b, where auxiliary information from the HST were used to train the GP model \citep{Gibson2012HST}.  Subsequently, we use their {\tt GeaPea} and {\tt Infer} modules\footnote{Written in {\tt Python} programming language and available freely from the following repository: \url{https://github.com/nealegibson}} for the definition of our GP and the implementation of Bayesian inference.

In the Bayesian framework, we write the likelihood function, $\mathcal{L}$, as a matrix equation in order to introduce off-diagonal elements to the covariance,
\begin{equation}
\log \mathcal{L} \left( \boldsymbol{r} | \boldsymbol{\mathsf{X}}, \theta, \phi \right) = -\frac{1}{2} \left( \boldsymbol{r}^T \boldsymbol{\Sigma}^{-1} \boldsymbol{r} + \log |\boldsymbol{\Sigma}| + N\log 2\pi \right)
\end{equation}
\noindent where $\boldsymbol{r}(=f_i-M_i)$ is the residual vector with $M$ being the transit model, $\boldsymbol{\mathsf{X}}$ is the $N \times K$ input parameter matrix ($K$ being the number of inputs and $N$ being the number of data points or flux measurements), $\theta$ and $\phi$ are collections of noise and transit model parameters respectively, and $\boldsymbol{\Sigma}$ represents the covariance matrix.  Ideally we would like to make every element of the covariance matrix a free parameter in our model, but the size of the matrix ($N \times N$) makes this rather impractical.  Hence we ``model" those elements using a covariance function, more commonly referred to as the \textit{kernel}.  With the aid of this kernel, the covariance matrix is written as $\Sigma_{ij} = k(x_i,x_j,\theta) + \delta_{ij}\sigma^2$, $x$ being an input of the kernel, $\delta$ the Kronecker delta and $\sigma^2$ the variance term, the last two of which ensure the addition of Poisson or white noise to the diagonal of the covariance matrix.

For the kernel, we use the Squared Exponential (SE)\footnote{Also known as the Radial Basis Function} kernel, which for a multi-dimensional parameter space ($K$), in its additive form, is written as,
\begin{equation}
k_{SE}(x_i,x_j,\theta) = \zeta \exp \left[ -\sum_{\alpha=1}^K \eta_\alpha \left( x_{\alpha,i}-x_{\alpha,j}\right)^2\right]
\end{equation}
\noindent where $\zeta$ is the maximum covariance and $\eta_\alpha$ are inverse scale parameters for all the input vectors $\boldsymbol{\mathsf{x}}$ (essentially the columns of the $\boldsymbol{\mathsf{X}}$ matrix).  This parameter determines the length of the ``wiggles" in the function.  We choose this form for the kernel as it is the de-facto default form for GPs as it is infinitely differentiable and easy to integrate against most functions.

Finally, in the definition of the analytical transit function, we use the quadratic limb-darkening \citep{Kopal1950} law to describe the centre to limb variations of brightness across the stellar disk.  We made this choice after the comparison of the Bayesian Information Criterion \citep[BIC;][]{Schwarz1978} values for models of higher complexity, such as the three parameter \citep{Sing2009} or the non-linear \cite{Claret2000} laws.

\subsubsection{Broadband}

We model the best two broadband differential transit light curves shown in Figure \ref{fig:Raw LCs}(c), which are produced by using the comparison stars 1 and 2 (top two), as the other light curves are either too noisy or suffer from large systematic effects.  These light curves are obtained by the integration of the relevant spectra within the domain that is shaded in grey in Figure \ref{fig:Raw Spec}.  In addition to the transit function, we also include a quadratic term (as a function of parallactic angle) in the description of the analytical transit model, to account for the \textbf{telescope rotation dependent} effects in the differential light curves.

To find the maximum posterior solution, $\mathcal{P}$, we need to optimise the Bayesian relation,
\begin{equation}\label{eq:log posterior}
\log \mathcal{P} \left( \theta, \phi | \boldsymbol{f},\boldsymbol{\mathsf{X}} \right) = \log \mathcal{L} \left( \boldsymbol{r} | \boldsymbol{\mathsf{X}}, \theta, \phi \right) + \log \mathrm{P} \left(\theta,\phi\right)
\end{equation}
\noindent which is true up to a constant term, where the final term on the right hand side of the equation is the prior probability assumed for the parameters of the complete model.  We generally use uninformative priors for the transit parameters, with the exception of the linear and quadratic limb darkening coefficients ($c_{1,2}$), the orbital period, $P$, and the eccentricity, $e$, according to:
\begin{equation}
\mathrm{P}(\phi) = \begin{cases}
	\mathcal{N}(	3.06785234,0) & \text{for } \phi \in [\text{P}],\\
	\mathcal{N}(	0,0) & \text{for } \phi \in [e],\\
	0 & \text{if $c_1 + c_2 > 1 $},\\
	1 & \text{otherwise}.
	\end{cases}
\end{equation}
\noindent which fixes the period and eccentricity of the orbit to what has previously been determined by \cite{Triaud2015}, and also ensures positive brightness across the stellar disk.  For the noise model parameters, we set the following prior probability distributions,
\begin{equation}
\mathrm{P}(\theta) = \begin{cases}
	\Gamma(1,1) & \text{for }\theta \in \left[\zeta , (\eta_1,...,\eta_K)\right],\\
	0 & \text{if $\sigma_w < 0 $},\\
	1 & \text{otherwise}.
	\end{cases}
\end{equation}
\noindent to ensure positive values for all the parameters, as well as using a gamma distribution\footnote{The probability density function for a gamma $\Gamma (k,\theta)$ distribution of a variable $x$ is given by $f(x,k,\theta) = x^{k-1}\exp (-x/\theta)/\theta^k$.} with shape and scale parameters both set to unity, $\Gamma(1,1)$, to encourage their values towards zero if the inputs that they represent are truly irrelevant in explaining the data \citep{Gibson2012GPs}.

For the starting values of the fitted parameters, we use the Nelder-Mead simplex algorithm \citep{Nelder1965} to find an initial set of optimal solutions for the transit and noise parameters.  To find the maximum posterior solution, we optimise the log posterior given in Equation \ref{eq:log posterior}.  We obtain these parameter posterior distributions by using the Monte-Carlo Markov-Chains method to explore the joint posterior probability distribution of our multivariate models \citep{Cameron-Collier2007,Winn2008,Gibson2012GPs}.  We run four independent MCMC simulations
with 100 walkers, of 100 000 iterations each. Once the chains are computed, we extract the marginalised posteriors for the free parameters to check for mutual convergence and for possible correlations among any pairs of parameters.  The transit parameters that are fitted for are mid-transit time ($T_0$), scaled semi-major axis ($a/R_\star$), relative planetary radius ($R_p/R_\star$), impact parameter ($b$), limb-darkening coefficients ($c_1~\&~c_2$), the three coefficients of the baseline model.  This model is a second order polynomial of the parallactic angle, \textit{q}, whose complexity is chosen using again the BIC selection rule.  This baseline model was initially chosen as a quadratic in time, which would describe low frequency variations due to the colour difference between the target and the comparison star.  However, this approach results in out of transit flux fit that is worse than using the parallactic angle, and additionally leads to an overestimation of planetary radius as compared to previous results from photometry in our wavelength region.  In addition to these, we also fit for the noise model parameters ($\zeta,~[\eta_1,...,\eta_K],~\sigma^2$).

\begin{figure}
\includegraphics[width=\linewidth]{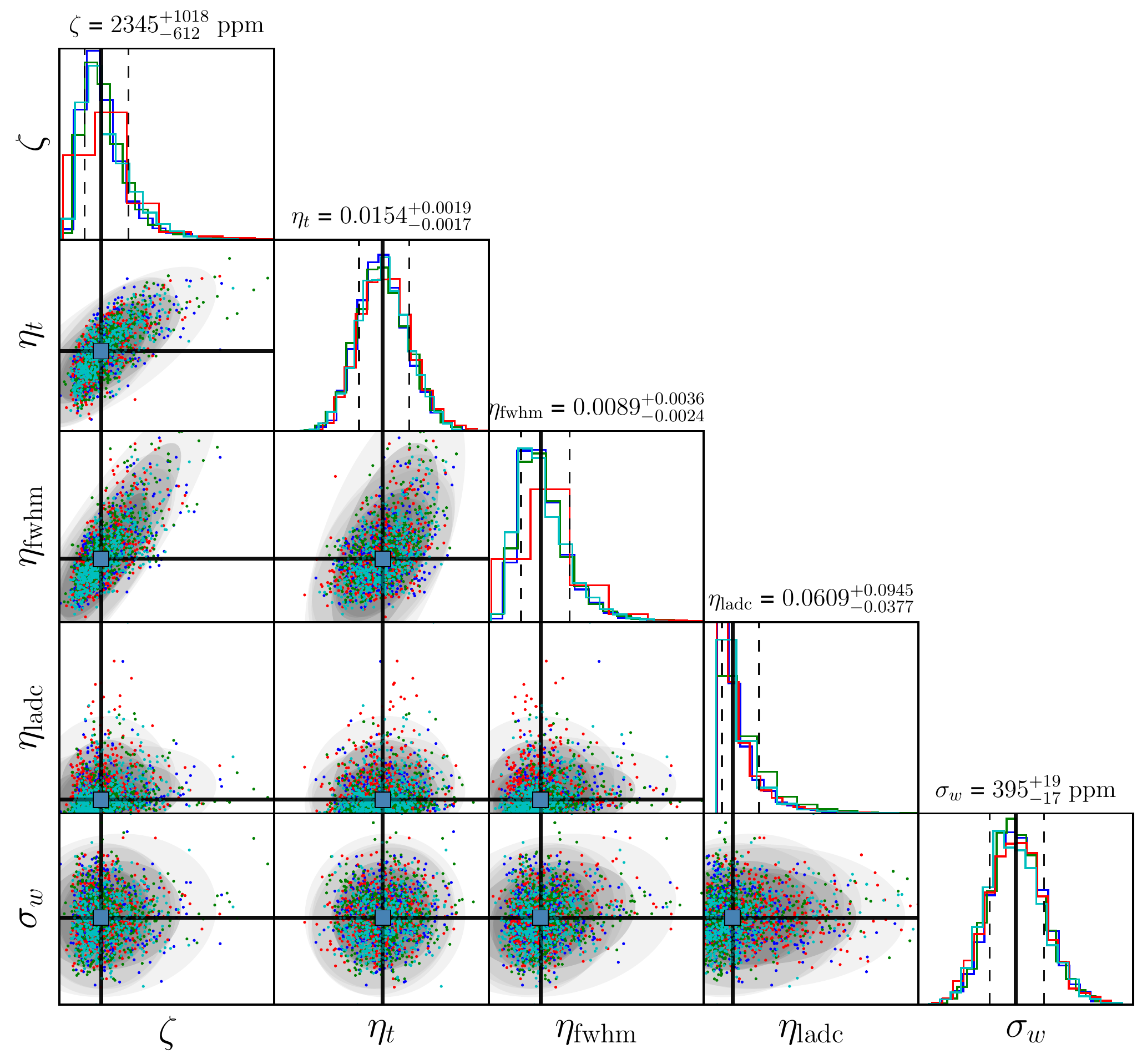}
\caption{Correlation plots for the noise model parameters, as well as their posterior distributions, shown for the broadband transit light of WASP-80b obtained relative to reference star 1.  The four colours correspond to the 4 independent MCMC simulations, each of 100 000 length.  For the scatter plots, we choose a random subset from each chain.  The final quoted parameter values are plotted as blue squares and their values, together with their associated errors, are given on top of each relevant posterior plot.}
\label{fig:noise pars}
\end{figure}

In order to decide what input parameters to include in the description of the GP model, as well as which of the two final broadband light curves suffer less from systematic effects, we model both of these broadband data series with our previously described model using up to three input parameters for the kernel.  We did not find any evidence of correlation with any other optical state parameters, to justify their initial inclusion into our model.  To compare the fitted models, we calculate their BIC values which includes a penalty term for addition of complexity to a model.  These calculated values are given in Table \ref{tab:model comp}, where their comparison indicates that our systematic model is best able to describe the broadband light curve relative to the comparison star 1, when \textit{time}, \textit{seeing} and the \textit{ladc inhomogeneity} are all used to model the covariance matrix.  However, it must be noted that there is some evidence against this model relative to the case that does not include the LADC variations ($\Delta$BIC$\le$6), but we choose to include this final parameter due to the small increase in the reduced chi-squared statistic ($\chi^2_\nu$).  The correlations and posterior distributions of the parameters of this noise model are given in Figure \ref{fig:noise pars}.
\begin{table}
  \centering
  \caption{Statistical comparison of our systematic noise model for various inputs, as well as the choice of reference star.  The highlighted value indicates the final choice of our data and model setup.}
    \begin{tabu} to 0.88\linewidth {X[4,l]
                                X[1,c]
                                X[1,c]
                                X[1,c]
                                X[1,c]}
    \addlinespace
    \hline \hline
    & \multicolumn{2}{c|}{Ref 1} & \multicolumn{2}{c}{Ref 2} \\
    \midrule
    GP Input parameters, $\eta_i$ & $\chi^2_\nu$  & \multicolumn{1}{c|}{BIC} & $\chi^2_\nu$  & BIC \\
    \hline
    $t$ & 1.85 & \multicolumn{1}{c|}{559} & 1.63 & 504 \\
    $t$, psf fwhm & 0.99 & \multicolumn{1}{c|}{335}& 1.35 & 430 \\
    $t$, F$_{\mathrm{ladc}}$ & 1.72 & \multicolumn{1}{c|}{546} & 1.65 & 513 \\
    $t$, psf fwhm, F$_{\mathrm{ladc}}$ & 1.00 & \multicolumn{1}{c|}{\textbf{341}} & 1.41 & 451 \\
    \hline
    \end{tabu}
    \label{tab:model comp}
\end{table}
From these posterior probability distributions we infer that our noise model input parameters are significantly relevant in describing the data, as the inverse length scale values, $\eta_i$, for all the inputs are $>3\sigma$ away from 0, with the exception of $\eta_{\mathrm{ladc}}~(\sim 2\sigma)$.

\begin{figure}
\includegraphics[width=\linewidth]{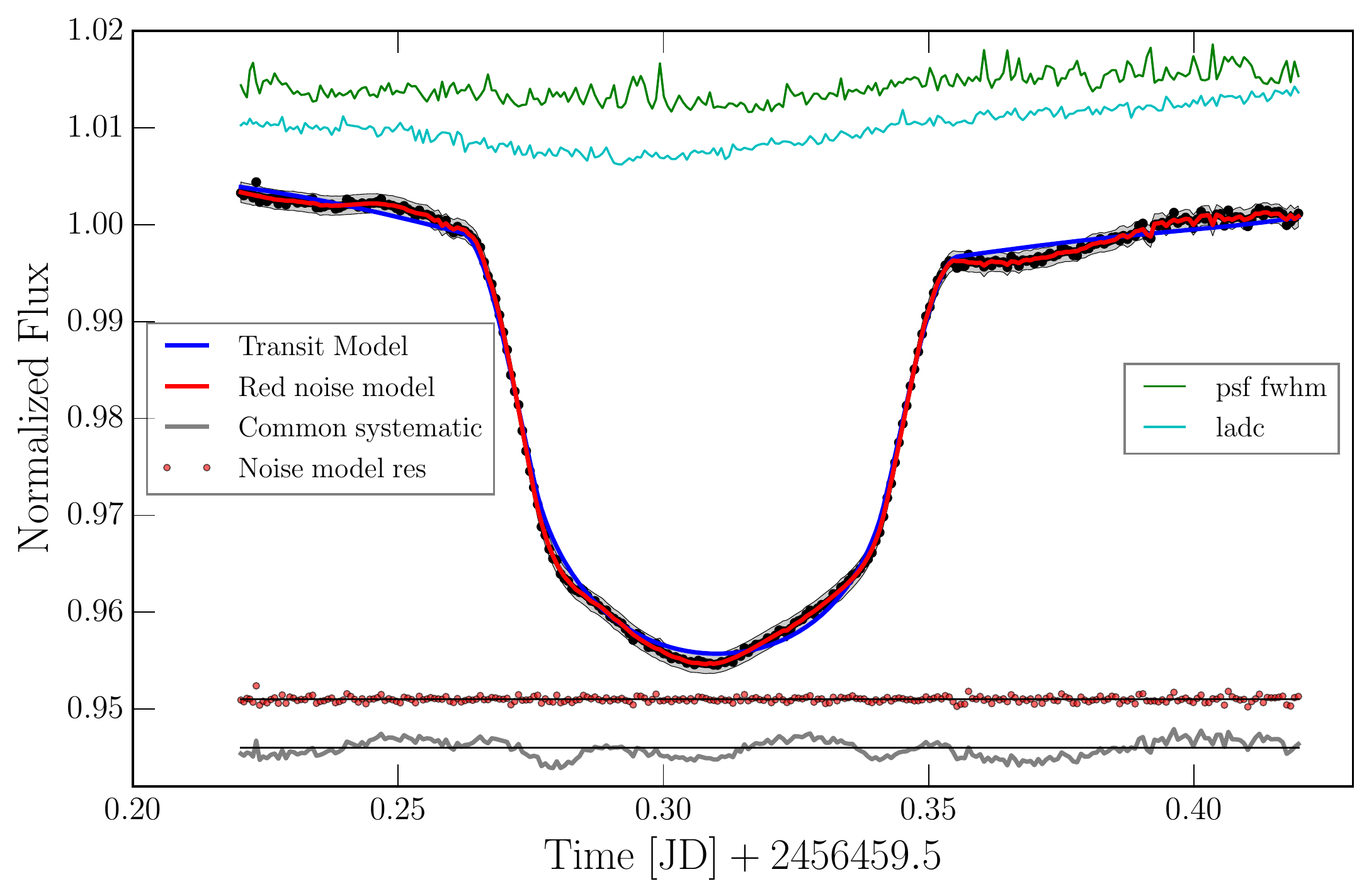}
\caption{Broadband transit light curve of WASP-80b.  The blue line is the best fit transit model and the red line is our systematic model with three inputs and the dark and grey areas are the 1- and 3-$\sigma$ confidences of that model, respectively.  The residuals of this model are shown as red dots below the light curve, shifted for clarity.  The grey line represents the common mode correction, which is essentially the residuals of the transit model.  The green and cyan lines at the top of the plot are the inputs of the covariance kernel, i.e. the PSF FWHM and the differential LADC discrepancy, respectively.}
\label{fig:broadband LC}
\end{figure}

The best fit noise model, as well as the corresponding transit function are given in Figure \ref{fig:broadband LC} (as red and blue lines respectively), for the broadband light curve obtained relative to reference star 1, as discussed previously.  The inferred transit and noise model parameters from this fit are given in Table \ref{tab:broadband results}, together with their 1-$\sigma$ uncertainties.
\begin{table}
\caption{The inferred transit parameters of WASP-80b from the analysis of MCMC simulations.}
\begin{tabular}{l l}
\hline \hline
Transit Parameter & Value \\
\hline
Mid-Transit, T$_0$, [JD] +2456459 & 0.80879 $\pm$ 7.3$\times$10$^{-5}$\\
Barycentric corrected T$_0$ [BJD$_{\mathrm{TDB}}$] & 2456459.809578 \\
Scaled semi-major axis, $a/R_\star$ & 12.0647 $\pm$ 0.0099\\
Relative planetary radius, $R_p/R_\star$ & 0.17386 $\pm$ 0.00030\\
Impact parameter, $b$ & 0.2325 $\pm$ 0.0087\\
Linear LD coefficient, $c_1$ & 0.491 $\pm$ 0.238\\
Quadratic LD coefficient, $c_2$ & 0.485 $\pm$ 0.198 \\
\hline
Noise Parameter & \\
\hline
Maximum covariance, $\zeta$ [ppm] & $2345^{+1018}_{~~~-612}$ \\
Time inverse scale parameter, $\eta_t$ & $0.0140\pm0.0018$\\
fwhm inverse scale parameter, $\eta_{\mathrm{fwhm}}$ & $0.0089^{+0.0036}_{~~~-0.0024}$ \\
ladc inverse scale parameter, $\eta_{\mathrm{ladc}}$ & $0.0609^{+0.0945}_{~~~-0.0377}$ \\
White noise, $\sigma$ [ppm] & $395\pm 18$\\
\hline
Derived parameter & \\
\hline
Planet radius, $R_p$ [$R_{\mathrm{jup}}$] & 0.99$^a$ $\pm$ 0.03\\
Orbital inclination, $i$ [$^{\circ}$] & 88.90 $\pm$ 0.06\\
\hline
\multicolumn{2}{l}{$^a$ \footnotesize Calculated using the stellar radius of $0.586\pm0.018R_{\odot}$}\\ 
\multicolumn{2}{l}{~~~from \citet{Triaud2015}.}
\end{tabular}
\label{tab:broadband results}
\end{table}
Our inferred transit parameters are in statistical agreement with those values reported by previous studies, covering a similar wavelength domain as our FORS2 observations \citep{Triaud2013,Triaud2015,Mancini2014,Fukui2014}.  We particularly use those values obtained by \citet{Mancini2014} as a benchmark due to their high precision light curves and the common wavelength coverage of their photometric light curves with our observations, namely Sloan \textit{i}' and \textit{z}'.

\subsection{Transmission spectrum}
\subsubsection{Unocculted spots}
The presence of active regions on the stellar surface facing the observer, be it spots or faculae, has a differential impact upon our spectroscopic light curves as it is a chromatic effect \citep{Csizmadia2013}.  Generally, this differential effect is more prominent in the near-UV wavelength domain, where the difference between the spectral density functions of the photosphere and the spot is more notable.  Regardless, we do still estimate possible spot-induced radius variations as a function of wavelength, to put any atmospheric interpretations into context.  Defining the dimming of the star, $1-f$, as the flux ratio of the star at an active phase (larger number of spots) to a relatively quiet phase, one estimates the spot filling factor, $\alpha$, from the relation

\begin{equation}
\alpha = \frac{1-f}{1-\left(\frac{T_\bullet}{T_\star}\right)^4} = \frac{1-f}{1-\beta ^4}
\end{equation}

\noindent where $T_\star$ is the stellar photospheric temperature, $T_\bullet$ is the spot temperature and $\beta$ is the fractional temperature difference of the spot and the surrounding photosphere.  Here we have taken the bolometric approximation to Planck's law.  \cite{Sarr1990} obtain spot temperatures $\sim750\pm150$K below the photospheric temperature, from excess absorption in the TiO bands of two K-dwarf stars, values that are further validated by \cite{Afram2015}.  With the stellar photosphere at $4143\pm93$K, and assuming a 10-20\% filling factor \citep{Andersen2015}, we estimate a 5-11\% flux dimming for the star.

Based on the above estimates, to correct for the presence of unocculted spots, we simulate light curves for each spectroscopic channel, using the parameter value obtained from that particular bin.  In addition, we include a spot on the stellar southern hemisphere (assuming the transit chord is in the north), with the above physical characteristics.  For each channel, only the spot contrast is adjusted, as it is the only wavelength dependent parameter.  Other spot parameters include two positional arguments and a further size argument, parameterised as spot radius.  We then model these simulated light curves with our previous model that does not include a spot.

From these simulations we obtain an average of $33\pm8$ppm discrepancy among the values of the relative planetary radii.  Therefore, we conclude that the wavelength-dependence of the planetary radius is unaffected by unocculted stellar spots.  We correct our eventual transmission spectrum for the presence of unocculted spots, although such correction terms are much smaller than the derived error bars of the relative planetary radii.

\subsubsection{Spectral light curves}
\begin{table}
\caption{Transmission spectrum and the two limb darkening coefficients from modelling the spectral light curves.}
\begin{tabular}{l c c c}
\hline \hline
Band (nm) & $R_p/R_\star$ & $c_1$ & $c_2$\\
\hline
748 $-$ 758 & 0.17563 $\pm$ 0.00100 & 0.528 $\pm$ 0.266 & 0.434 $\pm$ 0.287 \\ 
758 $-$ 768 & 0.17680 $\pm$ 0.00099 & 0.538 $\pm$ 0.258 & 0.438 $\pm$ 0.280 \\ 
768 $-$ 778 & 0.17747 $\pm$ 0.00063 & 0.521 $\pm$ 0.352 & 0.430 $\pm$ 0.297 \\ 
778 $-$ 788 & 0.17556 $\pm$ 0.00057 & 0.550 $\pm$ 0.213 & 0.442 $\pm$ 0.231 \\ 
788 $-$ 798 & 0.17558 $\pm$ 0.00060 & 0.542 $\pm$ 0.221 & 0.442 $\pm$ 0.260 \\ 
798 $-$ 808 & 0.17557 $\pm$ 0.00054 & 0.520 $\pm$ 0.361 & 0.430 $\pm$ 0.030 \\ 
808 $-$ 818 & 0.17479 $\pm$ 0.00048 & 0.520 $\pm$ 0.377 & 0.430 $\pm$ 0.029 \\ 
818 $-$ 828 & 0.17524 $\pm$ 0.00068 & 0.522 $\pm$ 0.233 & 0.430 $\pm$ 0.275 \\ 
828 $-$ 838 & 0.17365 $\pm$ 0.00063 & 0.525 $\pm$ 0.232 & 0.435 $\pm$ 0.270 \\ 
838 $-$ 848 & 0.17474 $\pm$ 0.00056 & 0.527 $\pm$ 0.217 & 0.431 $\pm$ 0.274 \\ 
848 $-$ 858 & 0.17348 $\pm$ 0.00055 & 0.522 $\pm$ 0.218 & 0.433 $\pm$ 0.283 \\ 
858 $-$ 868 & 0.17319 $\pm$ 0.00062 & 0.522 $\pm$ 0.225 & 0.431 $\pm$ 0.274 \\ 
868 $-$ 878 & 0.17057 $\pm$ 0.00070 & 0.520 $\pm$ 0.311 & 0.430 $\pm$ 0.315 \\ 
878 $-$ 888 & 0.17141 $\pm$ 0.00061 & 0.525 $\pm$ 0.219 & 0.435 $\pm$ 0.276 \\ 
888 $-$ 898 & 0.17338 $\pm$ 0.00066 & 0.519 $\pm$ 0.236 & 0.430 $\pm$ 0.281 \\ 
898 $-$ 908 & 0.17102 $\pm$ 0.00062 & 0.516 $\pm$ 0.236 & 0.432 $\pm$ 0.276 \\ 
908 $-$ 918 & 0.17207 $\pm$ 0.00055 & 0.520 $\pm$ 0.030 & 0.430 $\pm$ 0.304 \\ 
918 $-$ 928 & 0.17144 $\pm$ 0.00069 & 0.512 $\pm$ 0.245 & 0.428 $\pm$ 0.271 \\ 
928 $-$ 938 & 0.17107 $\pm$ 0.00070 & 0.501 $\pm$ 0.235 & 0.420 $\pm$ 0.280 \\ 
938 $-$ 948 & 0.17178 $\pm$ 0.00071 & 0.520 $\pm$ 0.165 & 0.430 $\pm$ 0.338 \\ 
948 $-$ 958 & 0.17250 $\pm$ 0.00083 & 0.520 $\pm$ 0.290 & 0.430 $\pm$ 0.315 \\ 
958 $-$ 968 & 0.16855 $\pm$ 0.00096 & 0.520 $\pm$ 0.294 & 0.430 $\pm$ 0.302 \\ 
968 $-$ 978 & 0.17086 $\pm$ 0.00074 & 0.520 $\pm$ 0.308 & 0.430 $\pm$ 0.296 \\  
\hline
\end{tabular}
\label{tab:spectro res}
\end{table}

\begin{figure*}
\includegraphics[width=\textwidth]{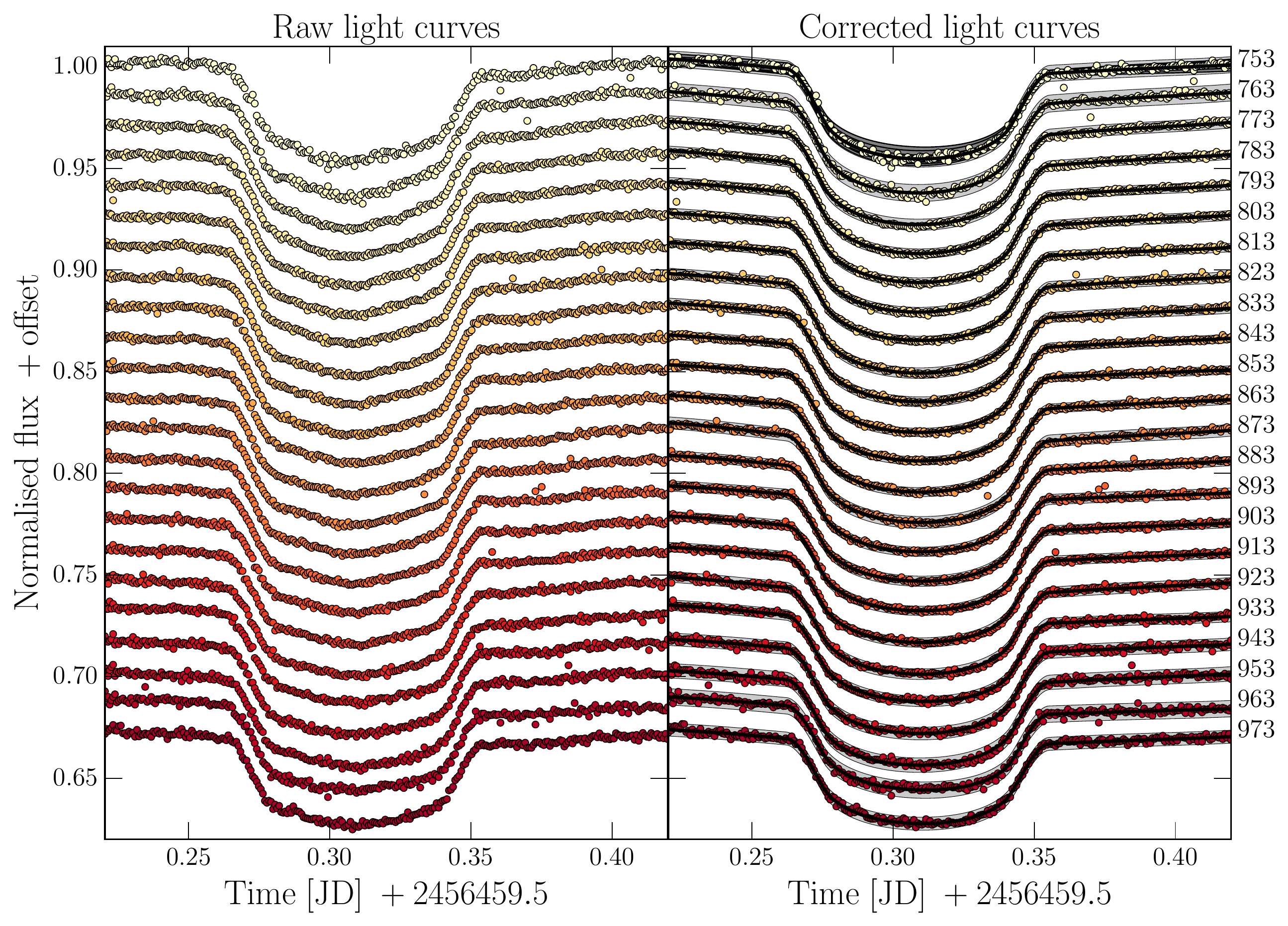}
\caption{Spectrophometric light curves of WASP-80b transit, obtained through narrowband integration of spectra from the FORS2+600z grism, normalised to out of transit flux and shifted for clarity.  The left panel shows the raw differential light curves, and the right panel presents the same light curves corrected for the common mode of systematics.  The central wavelength of each channel is given to the right of each light curve, with bin size of 100\AA.  Our best fit transit model for each channel is given as a solid black line, with the 3$\sigma$ confidence of each fit shaded as light grey.}
\label{fig:spectro LCs}
\end{figure*}

To obtain the wavelength dependent variations of the planetary radius, i.e. the \textit{transmission spectrum}, the spectra are integrated within smaller, 100\AA~bins, that are shown as dashed lines in Figure \ref{fig:Raw Spec}.  This bandwidth determines the resolution of the final transmission spectrum and is chosen through a statistical analysis of stellar spectra \citep{Sedaghati2016}.  This bandwidth is similar to most previous transmission spectroscopy studies performed with the FORS2 instrument \citep{Bean2011,Lendl2016,Sedaghati2016,Gibson2017} and those initial raw so-called spectrophotometric light curves are shown in the left panel of Figure \ref{fig:spectro LCs}.  As one looks for relative variations of transit depth in transmission spectroscopy, these spectral light curves are corrected for the common mode systematics that are present across all the channels \citep[e.g.][]{Lendl2016,Nikolov2016,Gibson2017}.  The common mode systematic correction is obtained through the division of broadband or white light curve by the analytical transit model that includes the baseline model as a quadratic function of the parallactic angle and no correlated noise component.  This correction sequence is shown as the grey line at the bottom of Figure \ref{fig:broadband LC}.  Subsequently, the spectral light curves are divided by this common correction term and modelled with the same systematic noise model as was done for the white light curve.  The corrected spectral light curves, together with the associated correlated noise models, are shown in the right panel of Figure \ref{fig:spectro LCs}.

\begin{figure*}
\includegraphics[width=\linewidth]{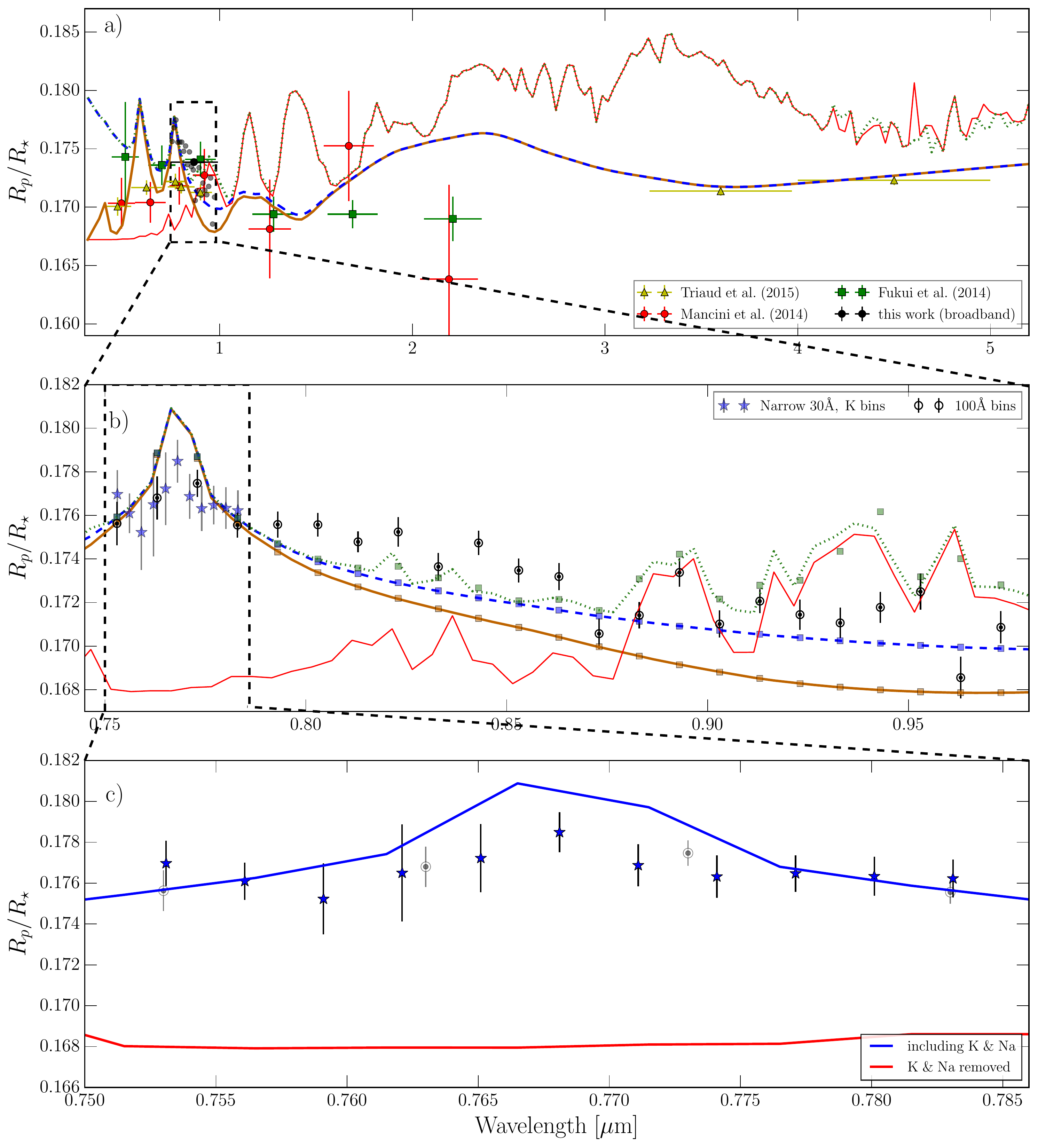}
\caption{\textit{(a)} Broadband transmission spectrum of WASP-80b stitched together from previous studies of this planet, as well as the broadband and spectroscopic points from our analysis.  The four overplotted atmospheric models are produced with the {\tt Exo-Transmit} code, at 800k equilibrium temperature, 0.1$\times$ solar metallicity.  The green dotted model includes the three molecules, H$_2$O, CO and CH$_4$, as well as H, He, Na and K, and the red without K and Na.  The blue dashed and brown solid lines are the same as the previous models, but with the molecular species removed, where in the brown model Rayleigh scattering has additionally been switched off.  \textit{(b)}  A zoom into the region, where our transmission spectrum of WASP-80b is obtained with the 600z grism of FORS2, produced with bins of 100\AA~width.  The plotted models are the same as above.  \textit{(c)} A further zoom into the region of absorption from the potassium line core and the wings.  The presence of potassium in the upper atmosphere is detected using the results from very narrow (30\AA) bins, where confirmation is made through a comparison with models that include potassium (\textit{solid blue, same as the dashed blue in the above plots}) and one that excludes it (\textit{solid red}).}
\label{fig:trans spec}
\end{figure*}

In modelling the spectral channels we set strict, informative priors on the values of model parameters that are wavelength independent (T$_0$, $a/R_\star$, $b$), as well as the three coefficients of the baseline model.  Since the limb darkening of the star is a chromatic aspect, the two coefficients of the law describing the centre to limb intensity variations are taken as free parameters in modelling the narrow-band light curves, with prior probabilities same as the broadband case.  One would also expect that seeing- and LADC-dependent systematics to be wavelength independent, which are removed by the common mode correction approach.  To confirm this, we allow their respective inverse scale parameters ($\eta_{\mathrm{fwhm}}$ and $\eta_{\mathrm{ladc}}$) to be free in our analysis, and check whether their posterior distributions approach zero for each channels, which in fact was the case for all light curves.   This procedure also ensures that uncertainties from any residual systematics are fully accounted for and their contributions to the final parameter error estimation are correctly propagated.  The transmission spectrum, obtained from this modelling process is given together with results from previous studies \citep{Mancini2014,Fukui2014,Triaud2015} in panel \textit{(a)} of Figure \ref{fig:trans spec}, and separately in panel \textit{(b)}.
And finally, the wavelength dependent parameters obtained from modelling the spectrophotometric light curves are given in Table \ref{tab:spectro res}.

Additionally we produce light curves from very narrow (30\AA) integration bins around the potassium doublet absorption region (0.766 \& 0.770 $\mu$m) for possible detection of K in the atmosphere of WASP-80b.  We modelled these light curves with the same approach as was done for the 100\AA~spectroscopic light curves, the results from which are shown in Figure \ref{fig:trans spec}c.

\section{Discussion}\label{sec:Discussion}

To interpret our observational results, we compare the transmission spectrum from the previous section to a variety of forward atmospheric models that explore a wide range of the parameter space, limited by abundance files provided by the atmospheric modelling code.  For this purpose, we use the open source, publicly available {\tt Exo-Transmit}\footnote{Available at the following repository:
\url{https://github.com/elizakempton/Exo_Transmit}} code \citep{Kempton2016} to model the planet size variations as a function of wavelength, using the opacity information provided \citep{Freedman2008,Freedman2014,Lupu2014}.  It must be stressed that we do not fit these spectra to the data and simply over-plot them for comparison.  \cite{Triaud2015} measured the brightness temperature of WASP-80b as $\sim900$K from \textit{Spitzer} photometry of occultations, with the equilibrium temperature calculated as $825\pm19$K.  Consequently, we calculated theoretical spectra for 700K, 800K and 900K equilibrium temperature T-P profiles, where a comparison of the reduced chi-squared statistic, $\chi^2_\nu$, as measured with the combined transmission spectrum (c.f. Figure \ref{fig:trans spec}a), suggests a significantly better fit for equilibrium temperature of 800K as compared to 700K, and marginally better with respect to 900K models.  This result is in agreement with the estimated equilibrium temperature of the exo-atmosphere.  Therefore, from here on, we only plot models calculated with the 800K equilibrium temperature T-P profile, and quote statistics based on this assumption.

\begin{figure}
\includegraphics[width=\linewidth]{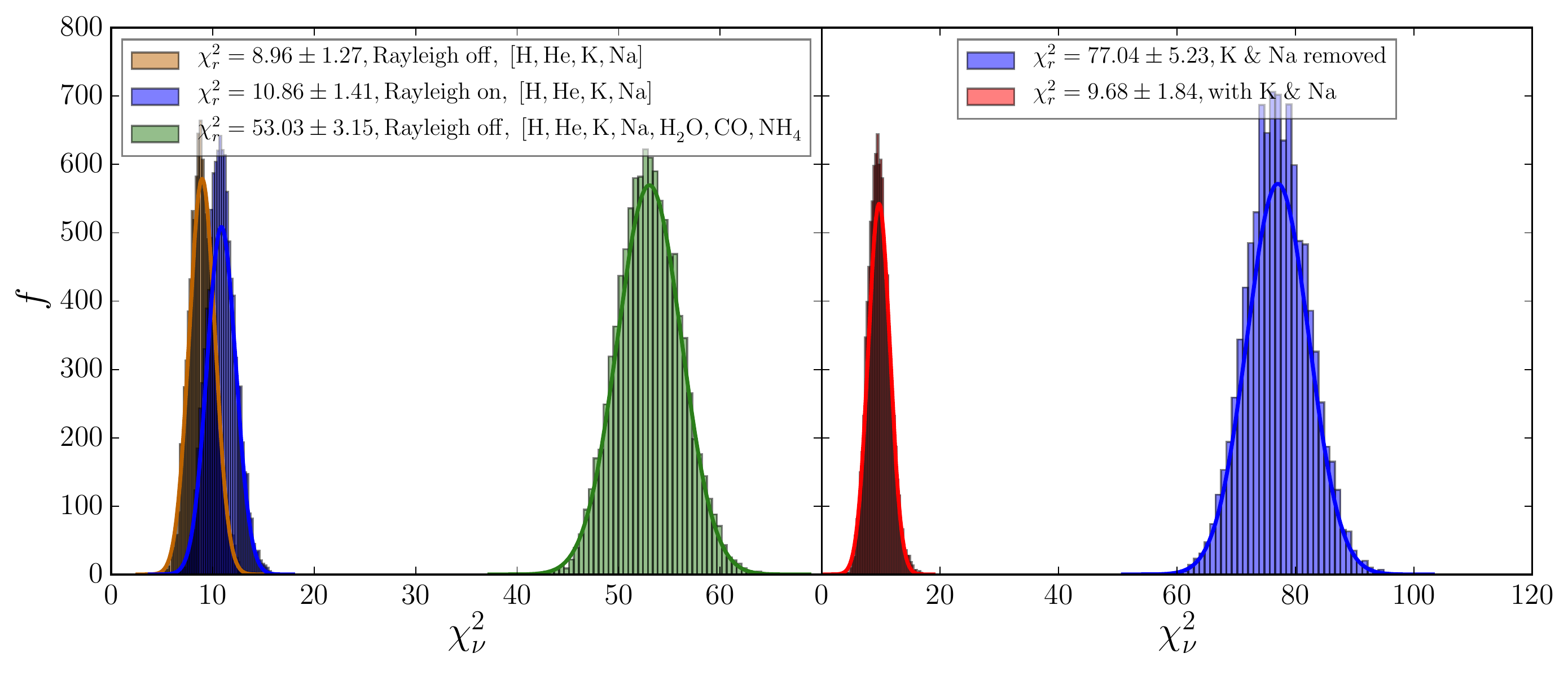}
\caption{\textit{(left)} $\chi^2_r$ distributions for three of the atmospheric models shown in Figure \ref{fig:trans spec}a, measured together with the data from broadband photometry.  This analysis significantly rules out additional absorption in the IR due to molecular species such as H$_2$O, CO or CH$_4$, and gives marginal preference to those that do not include a Rayleigh scattering signature.  (\textit{right}) Same analysis, but performed for the two atmospheric models in Figure \ref{fig:trans spec}c, where we significantly ($\gg 5\sigma$) detect the presence of potassium in the atmosphere of WASP-80b.}
\label{fig:atmos stat}
\end{figure}

Comparison to theoretically calculated models indicates a solar or sub-solar (1 or 0.1) metallicity for the atmospheric composition of this planet, with larger values (5$-$1000$\times$) yielding significantly larger $\chi^2_\nu$ statistics.  Furthermore, for all the models evaluated, we consider the case where condensation and rain out of molecules occurs.  Again the comparison significantly suggests no evidence for rain out and therefore we compute the spectra with only gas phase chemistry accounted for.  The complete transmission spectrum (Figure \ref{fig:trans spec}a) rules out the presence of absorption from H$_2$O, CO or NH$_4$ in the atmosphere, all of which have been included in the \textit{green dashed} and \textit{solid red} models, and removed from the other two.  We rule out their presence at $\gg5\sigma$ significance, calculated through the bootstrap of transmission spectrum values and their associated errors (the estimated $\chi^2_r$ distributions are shown in the left panel of Figure \ref{fig:atmos stat}).  This result is achieved mainly through the precise photometry of \textit{Spitzer} in the IR and GROND in the NIR.  Additionally, our analysis gives marginal preference to models without a Rayleigh scattering signature, although this conclusion is not statistically significant ($\sim 2\sigma$).
As mentioned earlier, our results also point to an atmosphere with sub-solar metallicity, although the exact level of it would need to be determined more precisely with retrieval models, once more spectroscopic data has been obtained over a larger wavelength range range -- a work that is outside the scope of this paper. We note that \citet{Triaud2015} find a metallicity for WASP-80b's atmosphere of $-0.13^{+0.15}_{-0.17}$, definitively leaving the possibility of sub-solar metallicity. 

The bluest part of the transmission spectrum is clearly dominated by the wings of the potassium line.  To investigate this further, we produce spectroscopic light curves using narrower, 30\AA, bins around the strong potassium doublet at $\sim0.786\mu$m.  The results from analysis of these light curves are given in Figure \ref{fig:trans spec}c, where comparison to atmospheric models with and without potassium included confirms the presence of this alkali metal at very high significance ($\gg 5\sigma$), in the upper atmosphere of WASP-80b, as was also previously done with FORS2 for the planet WASP-17b \citep{Sedaghati2016}.  This detection is again made through an estimation of the $\chi^2_r$ statistics from the bootstrap of the residuals, the results for which are shown in the right panel of Figure \ref{fig:atmos stat}.  High resolution spectroscopy of this target during and out of transit \citep[e.g.][]{Wyttenbach2015} will provide an alternative method for validation of this detection.

\section{Conclusions}\label{sec:Conclusions}
We report a ground-based red transmission spectrum of the exoplanet WASP-80b with the FORS2 instrument at ESO's VLT.  The instrument was used in the MXU mode to obtain simultaneous spectra of the target and the comparison stars with the 600z grism, covering a common wavelength domain of $\sim 0.75-1.05~\mu m$.  We successfully model the transit light curve with a Gaussian Process model with optical state parameters, such as seeing and the LADC inhomogeneities, as inputs.  These parameters are selected carefully through the analysis of their correlations with the light curve residuals.  We apply the standard common mode systematic correction to ``clean" the spectroscopic light curves, which are modelled with the same method as the broadband approach and different prior probabilities of some parameters.  Subsequently, the transmission spectrum is plotted, which describes the colour-dependent variations of the planetary radius, as is expected for an extended atmosphere.  An initial analysis of our results together with previously obtained transmission spectra of this planet, points to an equilibrium temperature in agreement with the value determined by \citep{Triaud2015}, as well as a sub-solar metallicity.  Furthermore, we rule out the presence of heavier molecules that lead to additional absorption in the IR, at high significance.  However, a word of caution here is that the analysis is based on only a few data points in the IR domain and more observations will be required to validate this conclusion.  The analysis of our transmission spectrum obtained with FORS2, points to excess absorption in the core and the extended pressure-broadened wing of the potassium doublet.  Further analysis of this pressure-broadening of the wings will result in an estimation of the cloudiness index for this planet, as proposed by \citet{Heng2016}.  The presence of potassium in the atmosphere of WASP-80b is detected through the modelling of additional narrowband (30\AA) transit light curves, where models including potassium lead to significantly ($\gg 5\sigma$) better fit statistics, as compared to those models where potassium is removed.  Further observations with other grisms of this instrument will be required to extend the wavelength coverage of the spectrum and enable comparison with more complex atmospheric models.  Only then, one could statistically search the atmospheric parameter space, using retrieval methods \citep{Madhu2009}, to infer a more detailed structure of this exo-atmosphere.

We conclude that the potential of exo-atmospheric observations from ground-based facilities is encouraging, once the sources of correlated noise are carefully analysed.  In other words, it is imperative to understand the instrument that is collecting the light, as it was recently and very elegantly demonstrated by \citet{Deming2017}, and to try to exhaust the search for possible sources of correlated noise.  This is generally more often the case for observations made from space observatories where systematic noise dominates, but nevertheless should also closely be considered when a claim of atmospheric detection from the ground is made.  It must be stressed that our detections are only made possible through the use of an instrument-dependent baseline model and a thorough analysis of the sources of systematic noise in the light curves.  Finally, FORS2 is the leading instrument in the world for performing such observations \citep{Bean2010,Bean2011,Sedaghati2015,Sedaghati2016,Lendl2016,Nikolov2016,Gibson2017}, specially after the recent upgrade of the atmospheric dispersion corrector prisms, which has reduced this aforementioned systematic noise in the transit light curves significantly.

\section*{Acknowledgments}
All the data, together with the calibration files, are publicly available and can be downloaded from ESO's science archive, under the programme ID 091.C-0377(C).  We also acknowledge the use of several {\tt python} modules, such as {\tt numpy, matplotlib, scipy, astropy, pyfits, PyAstronomy}, as well as those mentioned in the text.  ES would like to acknowledge funding and support from ESO under the studentship programme.  LD acknowledges support from the Gruber Foundation Fellowship.  MG is an F.R.S.-FNRS research associate.  SzCs thanks the Hungarian National Research, Development and Innovation Office, for the NKFIH-OTKA K113117 grant.




\bibliographystyle{mnras}
\bibliography{WASP80}








\bsp	
\label{lastpage}
\end{document}